\documentclass[twocolumn,showpacs,prl,superscriptaddress,floatfix]{revtex4}
\usepackage{amsmath,amssymb}
\usepackage{graphicx}

\def\(({\left(}
\def\)){\right)}                       
\def\[[{\left[}
\def\]]{\right]}

\newcommand{\be}{\begin{equation}}
\newcommand{\ee}{\end{equation}}
\newcommand{\bea}{\begin{eqnarray}}
\newcommand{\eea}{\end{eqnarray}}

\begin{document}

\title{Locked constraint satisfaction problems}

\author{Lenka Zdeborov\'a}
\affiliation{ Universit\'e Paris-Sud, LPTMS, UMR8626,  B\^{a}t.~100, Universit\'e
Paris-Sud 91405 Orsay cedex}
\affiliation{CNRS, LPTMS, UMR8626, B\^{a}t.~100, Universit\'e Paris-Sud 91405 Orsay cedex}

\author{Marc M\'ezard}
\affiliation{ Universit\'e Paris-Sud, LPTMS, UMR8626,  B\^{a}t.~100, Universit\'e
Paris-Sud 91405 Orsay cedex}
\affiliation{CNRS, LPTMS, UMR8626, B\^{a}t.~100, Universit\'e Paris-Sud 91405 Orsay cedex}

\begin{abstract}
  We introduce and study the random `locked' constraint satisfaction problems.
  When increasing the density of constraints, they display a broad `clustered'
  phase in which the space of solutions is divided into many {\it isolated }
  points. While the phase diagram can be found easily, these problems,
  in their clustered phase, are extremely hard from the algorithmic point of
  view:  the best known algorithms all fail to find solutions. We thus propose
  new benchmarks of really hard optimization problems and provide insight
  into the origin of their typical hardness.
\end{abstract}

\pacs{89.70.Eg,75.10.Nr,64.70.P-}
\date{\today}
\maketitle

Constraint satisfaction problems (CSPs) are one of the main building blocks of
complex systems studied in computer science, information theory and
statistical physics. Their wide range of applicability arises from their very
general nature: given a set of $N$ discrete variables subject to $M$
constraints, the CSP consists in deciding whether there exists an assignment
of variables which satisfies simultaneously all the constraints. In computer
science CSPs are at the core of computational complexity studies: the
satisfiability of boolean formulas is the canonical example of an
intrinsically hard, NP-complete, problem~\cite{Cook71}. In information theory
error correcting codes also rely on CSPs. The transmitted information is
encoded into a codeword satisfying a set of constraints, so that information
may be retrieved after transmission through a noisy channel, using the
knowledge of the constraints. Many other practical
problems in scheduling a collection of tasks or in hardware and software verification and testing are viewed as CSPs. In statistical physics the interest in CSPs
stems from their close relation with theory of spin glasses. Answering if
frustration is avoidable in a system is a first, and sometimes highly
nontrivial, step in understanding its low temperature behavior.

Methods of statistical physics provide powerful tools to study statistical
properties of CSPs \cite{MezardParisi01,MezardParisi03}. The mean field
approach is known to be exact if the underlying graph of constraints~\cite{KschischangFrey01} is either fully connected or locally
tree-like. It also has algorithmic, and practical, consequences: in contrast with the usual
situation in physics, CSPs on a locally tree-like graph are used in practice,
for instance in low density parity check codes~\cite{Gallager62}, which are
among the best error-correcting codes around.
 
Many CSPs are NP-complete. Nevertheless, large classes of instances can be
easy to solve. It is one of the main goals of theoretical computer science to
understand why some instances are harder than others, where the hardness comes
from and how to avoid it, beat it or use it. The random K-satisfiability
($K$-SAT) problem where clauses are chosen uniformly at random between all
possible ones has played a prominent role in approaching this goal. In random
$K$-SAT there exists a sharp satisfiability threshold. This is a phase
transition point separating a 'SAT' phase with low density of constraints
where instances are almost always satisfiable, from an 'UNSAT' phase where, with high
probability, there is no solution to the CSP
\cite{KirkpatrickSelman94,Friedgut99}. The hardest instances lie near to this
threshold \cite{CheesemanKanefsky91,MitchellSelman92}. The main insight came
from statistical physics studies
\cite{MonassonZecchina99,BiroliMonasson00,MezardParisi02,MezardZecchina02,KrzakalaMontanari06,ZdeborovaKrzakala07,MontanariRicci08}
which allow to describe the structure of the space of solution of the random
$K$-SAT problem. The most interesting result is the existence of an
intermediate ``clustered'' phase, just below the SAT-UNSAT threshold, where the
space of solutions splits into well separated clusters. A major open question
consists in understanding if and how the existence of clusters makes the
problem harder. The survey propagation algorithm, which explicitly takes into
account the clusters, is the best known solver very close to the SAT-UNSAT threshold~\cite{MezardZecchina02}, but some local search algorithms also perform well
inside the clustered phase \cite{ArdeliusAurell06,KrzakalaKurchan07}. Another
proposition, put forward in~\cite{ZdeborovaKrzakala07}, is that solutions in clusters with frozen variables, taking the same value in the whole cluster, are hard to find. 
It was shown in~\cite{DallAstaRamezanpour08} that, even if solutions belonging to clusters without frozen variables are exponentially rare, some message passing algorithms may be able to find them.

In this letter we introduce and study a broad class of CSPs which are
extremely frozen problems: all the clusters consists of a single
configuration, thus all the variables are frozen in every cluster. We show
that these problems are extremely difficult from an algorithmic point of view:
all the best known algorithms fail to solve them in this clustered phase. At the
same time the description of their phase diagram can be carried out in details
with relatively simple statistical physics methods.

\paragraph{Definition --} We define an {\it occupation CSP} over $N$ binary
variables, $s_1,\dots,s_N\in \{0,1\}$ as follows:  each constraint $a$
connects to $K$ randomly chosen variables, and its status depends on the sum
$r$ of these variables. The constraint is characterized by a $(K+1)$
component vector $A=(A_0A_1\cdots A_K)$, with $A_r\in\{0,1\}$: it is satisfied
if and only if $A_{r}=1$. We shall study here homogeneous models in which
all constraints connect to the same number~$K$ of variables, and are
characterized by the same vector~$A$. According to~\cite{Schaefer78} the 
occupation CSPs are NP-complete if $K>2$, $A_0=A_K=0$ and $A$ is not a parity check. 
The {\it locked occupation problems}
(LOP) are occupation CSPs satisfying two conditions: (a) $\forall i=0,\dots,K-1$ the
product $A_i A_{i+1}=0$, (b) all variables are present in at least two
constraints. Simple examples of LOPs are positive 1-in-3 satisfiability \cite{GareyJohnson79}, $A=0100$, or parity checks  \cite{Gallager62}, $A=01010$, on graphs
without leaves. In order to go from one solution (satisfying assignment) of a
LOP to another one, it is necessary to flip at least a closed loop of
variables in the factor graph representation of \cite{KschischangFrey01}.
This stays at the root of the crucial property that clusters are point-like
and separated by an extensive distance when the density of constraints is large enough (above $l_d$). 

In order to fully characterize a random LOP ensemble, one needs to define the
degree distribution of variables. We will study here two ensembles. The
regular ensemble, where every variable appears in exactly $L$ constraints, and
the truncated Poisson ensemble with degree distribution $Q(0)=Q(1)=0,
Q(l)=e^{-c} c^l /l! [1-(1+c)e^{-c}], l\ge 2$ and average connectivity
$\overline l = c (1-e^{-c})/[1-(1+c)e^{-c}]$.

\paragraph{Phase diagram --} Denoting by $a,b,\dots$ the indices of
constraints and $i,j,\dots$ those of variables, the belief propagation (BP) equations
\cite{YedidiaFreeman03} are given by:
\bea
\psi_{s_i}^{a\to i} &=& \frac{1}{Z^{a\to i}} \sum_{\{s_j\}} \delta(A_{s_i+\sum_{j} s_j} -1)\prod_{j\in {\partial a}-i} \chi_{s_j}^{j\to a} \, , \label{BP_1} \\
\chi_{s_j}^{j\to a} &=& \frac{1}{Z^{j\to a}} \prod_{b\in {\partial j}-a}
\psi_{s_j}^{b\to j}\, , \label{BP_2} \eea
 where $\partial a$ are all the
variables appearing in constraint $a$, and $\partial i$ all the constraints
in which variable $i$ appears. $\chi_{s_j}^{j\to a}$ is the probability that spin $j$ takes
value $s_j$ when $a$ was removed from the graph, and $Z$ are
normalization constants. The BP entropy (the logarithm of number of configuration satisfying all constraints,
divided by $N$) is
 \be
 s=\frac{1}{N}\sum_a \log{(Z^{a+\partial a})} - \frac{1}{N}
\sum_i (l_i-1) \log{(Z^i)}\, , \label{ent}
 \ee
 where:
 \bea
Z^{a+\partial a}&=& \sum_{\{s_i\}} \delta(A_{\sum_i s_i}-1) \prod_{i\in {\partial a}} \left( \prod_{b\in {\partial i}-a} \psi_{s_i}^{b\to i} \right) , \label{Za}\\
Z^i&=& \prod_{a\in {\partial i}} \psi_{0}^{a\to i}+ \prod_{a\in {\partial i}}
\psi_{1}^{a\to i}\, . \label{Zi} 
\eea
In order to find a fixed point of
eqs.~(\ref{BP_1}-\ref{BP_2}) and compute the quenched average of the entropy we use the
population dynamics technique \cite{MezardParisi01}, with population sizes 
of order $10^4$ to $10^5$. It turns out that this procedure always converges to the same fixed point.

The phase diagram of LOPs is much simpler to analyze than
the one of general CSPs, and can be deduced purely from the BP analysis. This
is due to the fact that, in the clustered phase, every cluster reduces to a
single isolated configuration. The survey propagation (SP) equations
\cite{MezardZecchina02} are then greatly simplified. Their iteration either
leads to a trivial fixed point, where every variable is in the so called
"joker" state \cite{BraunsteinMezard02}, or to a fixed point where no variable
is in the "joker" state. In this second case the SP equations reduce to the BP
eqs.~(\ref{BP_1}-\ref{BP_2}), and the complexity function (logarithm of number
of clusters) is equal to the entropy~(\ref{ent}), in agreement with the
point-like nature of clusters. The clustered phase is then identified from the
iterative stability of this second, non-trivial fixed point. 
It is iteratively stable when the average connectivity is above a threshold:  $\overline l > l_{\rm d}$, while the
regime $\overline l < l_{\rm d}$ corresponds to a 'liquid' phase. The
intuitive difference between the two phases is that in the clustered phase one
has to flip an extensive number of variables to go from one solution to
another, while in the liquid phase the addition of any infinitesimal temperature
is enough to be able to connect all solutions. 

The satisfiability threshold $l_{\rm s}$ is defined as follows: If the average
connectivity is $\overline l<l_{\rm s}$ then a satisfying assignment almost
surely exists (in $N \to \infty$), and if $\overline l>l_{\rm s}$ then there
is almost surely no satisfying configuration. In LOPs we can find $l_{\rm s}$
as the average connectivity at which the RS entropy (\ref{ent}) becomes zero.
Table.~\ref{tab} gives the values of clustering and satisfiability thresholds
for the non-trivial LOPs with $K\le 5$.

\begin{table}[!ht]
\begin{tabular}{|l|l||l|l|l|} \hline
\, A       & name       & $L_{\rm s}$\, &\, $l_{\rm d}$&\, $l_{\rm s}$\\ \hline \hline
0100    & 1-in-3     & 3    & 2.256(3)\,& 2.368(4)\, \\ \hline 
01000   & 1-in-4     & 3    & 2.442(3)  & 2.657(4)   \\ \hline 
00100*  & 2-in-4     & 3    & 2.513     & 2.827      \\ \hline 
01010*  & odd 4-PC   & 4    & 2.856     & 4          \\ \hline 
010000  & 1-in-5     & 3    & 2.594(3)  & 2.901(6)   \\ \hline 
001000  & 2-in-5     & 4    & 2.690(3)  & 3.180(6)   \\ \hline 
010100  & 1-or-3-in-5& 5    & 3.068(3)  & 4.724(6)   \\ \hline 
010010\, & 1-or-4-in-5& 4    & 2.408(3)  & 3.155(6)   \\ \hline 
\end{tabular}  
\caption{\label{tab} The clustering $l_{\rm d}$ and satisfiability $l_{\rm s}$ thresholds in the
 locked occupation problems for $K\le 5$ in the truncated Poisson ensemble. In the regular 
ensemble $L_{\rm s}$ is the first unsatisfiable or critical connectivity, the first clustered 
case is $L_{\rm d}=3$. The error bars originate in the statistical nature of the population 
dynamics technique. Symmetric LOPs where the satisfiability threshold can be computed analytically are indicated by~*.
}
\end{table}
When a LOP is symmetric, i.e., $A_r=A_{K-r}$ for all $r=0,\dots,K$, and this
$0-1$ symmetry is not spontaneously broken, the satisfiability threshold can
be computed rigorously using the 1st and the 2nd moment methods: The annealed
entropy $\langle Z \rangle \equiv \exp{(N s_{\rm ann})}$ is:
\be 
s_{\rm ann}(\overline l)=\log{2}+\frac{\overline l}{K}
\log{\left[2^{-K}\sum_{r=0}^K \delta(A_r-1) {K\choose r}\right] }\label{s_ann}
\, . 
\ee
By computing the second moment $\langle Z^2 \rangle$ and using the Chebyshev's
inequality, as in \cite{MezardRicci03,CoccoDubois03}, we have shown that the
annealed entropy is equal to the typical one, thus the satisfiability
threshold $l_{\rm s}$ is given by $s_{\rm ann}(l_{\rm s})=0$. Examples of
LOPs for which this works are the parity checks $A=01010$, as well as $A=00100, 0001000,
0010100$, etc. Note that for instance $A=010010$ does not belong to this class
because its $0,1$ symmetry is spontaneously broken.

\paragraph{Algorithms --} We attempt to find solutions to LOPs in their
satisfiable phase using three algorithms which are among the best for hard
random instances of the K-satisfiability problem: belief propagation decimation
(BPd) \cite{KrzakalaMontanari06} (which is the same as survey propagation
\cite{MezardZecchina02} in LOPs),  stochastic local search
(SLS) \cite{SelmanKautz94},  and reinforced belief
propagation (rBP) \cite{ChavasFurtlehner05}. 

In BPd one uses the knowledge of marginal variable probabilities from BP
equations in order to identify the most biased variable, fix it to its most
probable value, and reduce the problem. In $K$-SAT the SP decimation (which in
LOPs is equivalent to BPd) has been shown to be very efficient, on very large
problems, even very near to the satisfiability threshold
\cite{MezardZecchina02}. However, in LOPs the BP decimation fails badly. For
example in the 1-or-3-in-5 SAT problem, on truncated Poisson graphs with
$M=2\cdot10^4$ constraints,
the probability of success is about 25\% at $\overline l =2$, and less than
5\% at already $\overline l =2.3$, way below the clustering threshold
$ l_d\simeq 3.07$.

Although we do not know how to analyze directly the BPd process, some mechanisms
explaining the failure of the decimation strategy can be understood using the
approach of \cite{MontanariRicci07}. The idea is to analyze a slightly simpler
decimation process, where the variable to be fixed is chosen uniformly at
random and its value is chosen according to its exact marginal probability,
which is assumed to be approximated by BP. The reduced formula after
$\theta N$ steps is equivalent to the reduced formula created by choosing a
solution uniformly at random and revealing a fraction~$\theta$ of its
variables. The number of variables which were either revealed or are directly
implied by the revealed ones is denoted $\Phi(\theta)$. The performance of
this `uniform' BP decimation can be understood from the shape of the
function $\Phi(\theta)$, which we have computed from the cavity
method.
\begin{figure}[!ht]
  \resizebox{\linewidth}{!}{\includegraphics{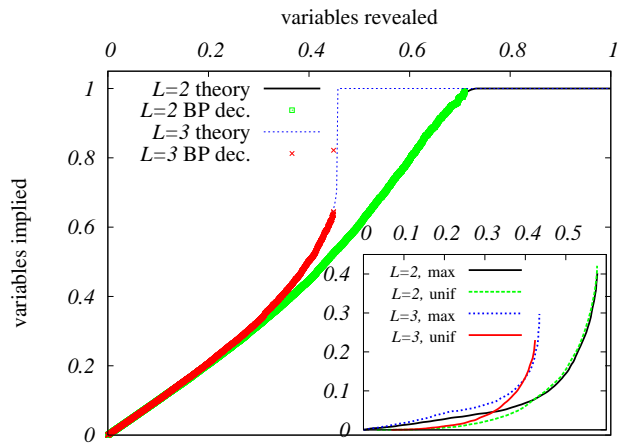}}
  \caption{\label{BP_anal} Uniform BP decimation in regular 1-or-3-in-5 SAT
    with $L=2$ and $L=3$: plot of $\Phi(\theta)$, as obtained analytically
    (lines) and from the uniform BP decimation (points): the two plots agree perfectly. For $L=3$
    the decimation fails because of avalanches at the discontinuity of
    $\Phi(\theta)$, for $L=2$ it fails when $\Phi(\theta)\to 1$ for
    $\theta<1$. Inset: Comparison between BPd and uniform BP
    decimation. The number of directly implied variables is plotted against
    number of variables which were free just before fixing them. The two methods
are very close, and they fail at about the same value of $\theta$.
\vspace{-0.3cm}
}
\end{figure}

In Fig.~\ref{BP_anal} we show that the theoretical curve $\Phi(\theta)$
agrees with numerical results in regular 1-or-3-in-5 SAT. At connectivity
$L=3$ the function has a discontinuity at $\theta_s\simeq 0.46$, thus after
fixing a fraction $\theta_s$ of variables an infinite avalanche of direct
implications follows and small errors in the BP estimation of marginals lead to a contradiction with high probability. At
connectivity $L=2$ the function $\Phi(\theta)\to 1$ at $\theta_1\simeq0.73$.
This means that if a fraction $\theta>\theta_1$ of variables in a random
solution is revealed the residual problem has only this single solution. Any
mistake in the previously fixed variables matters and causes a contradiction.
In all the LOPs we have studied, $\Phi(\theta)$ has one these two fatal
properties. The inset of Fig.~\ref{BP_anal} shows that, in LOPs, there is not
much difference in the behaviors of BPd and this uniform BP decimation.

\begin{figure}[!ht]
 \resizebox{\linewidth}{!}{\includegraphics{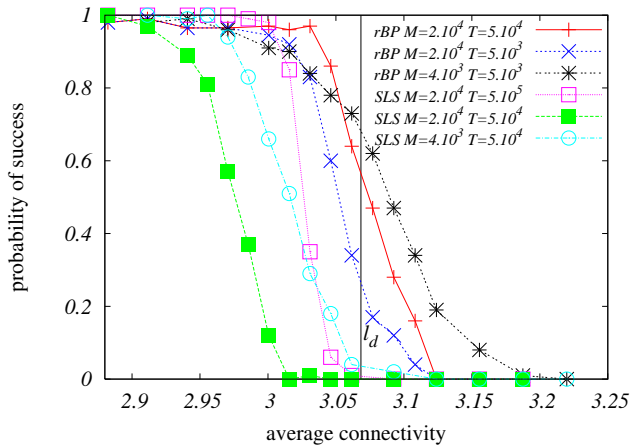}}
 \caption{\label{BP_SLS} Performance of reinforced BP and stochastic local
   search for 1-or-3-in-5 SAT with $M$ constraints. The fraction of
   successful runs is plotted against the average connectivity~$\overline l$.
   The clustering threshold $l_{\rm d}$ is marked, and the satisfiability
   transition is at $l_{\rm s}=4.72$. The maximal numbers of steps per variable $T$ are
   chosen such that the running times of rBP and SLS are comparable. 
\vspace{-0.3cm}
   }
\end{figure}

Stochastic local search (SLS) algorithms exist in many different versions and
are used in most practical cases where the exhaustive search is too time
consuming. The main idea of the family of algorithms is to perform a random walk in
configurational space, trying to minimize the the number of
unsatisfied constraints. In the implementation of \cite{ArdeliusAurell06},
a variable which belongs to at least one unsatisfied constraint is chosen
randomly. If flipping this variable does not increase the energy, the flip is
accepted. If it increases the energy, the flip is accepted with probability
$p$. This is repeated until either one finds a solution, or the number of steps per variable exceeds $T$. 
The parameter $p$ must be optimized. In Fig.~\ref{BP_SLS} we plot the
fraction of successful runs for the 1-or-3-in-5 SAT with $M$
constraints and $p=0.00003$.
Even with the largest value of $T$ we have not been able to solve instances with average connectivity
larger than $3.05$.

The belief propagation reinforcement (rBP) was originally introduced in
\cite{ChavasFurtlehner05}. The main idea is to add an external field
$\mu_{s_i}^i$ which biases the variable $i$ in the direction of the marginal
probability computed from the BP messages. This modifies BP eq.~(\ref{BP_2})
to $\psi_{s_i}^{i\to a} = \mu_{s_i}^i\prod_{b\in {\partial i}-a}
\psi_{s_i}^{b\to i}/Z^{i\to a} $. The algorithm then works as follows: Iterate
the BP equations $n$-times. Update all the external fields: If $\xi^i_1<\xi^i_0$
set $\mu_{1}^i=\pi^{l_i}, \mu_{0}^i=(1-\pi)^{l_i}$, otherwise set $\mu_{1}^i=(1-\pi)^{l_i}, \mu_{0}^i=\pi^{l_i}$, where $\xi^i_{s_i}=(\mu^i_{s_i})^{1/(l_i-1)} \prod_{a\in {\partial i}} \psi_{1}^{a\to i}$. At each iteration one checks if the most
probable configuration, given by $s_i=0$ if $\mu^i_{0}>\mu^i_{1}$ and $s_i=1$
otherwise, is a solution. If it is not one iterates at maximum $T$ times. 
We chose $n=2$ and
optimized the value of $\pi$. In Fig.~\ref{BP_SLS} we plot the fraction of
successful runs for the 1-or-3-in-5 SAT with $\pi=0.42$ for $2.8<\overline l<3$
and $\pi=0.43$ for $3\le \overline l<3.2$. The performance is marginally better than SLS,
but again one cannot penetrate into the clustered phase. 

We have observed the same behavior for all LOPs we studied: the clustering
transition point $l_d$ seems to be a boundary beyond which all these
three algorithms fail. As shown in Table~\ref{tab}, this point can be
very far from the SAT-UNSAT transition $l_s$, meaning that there is a
broad range of instances where known algorithms are totally inefficient. The parity check problems
are the exception as they can be solved with linear programming algorithms.

\paragraph{Conclusions --} LOPs make a broad class of extremely hard constraint
satisfaction problems. Their phase diagram is
simple: the set of satisfiable configurations becomes clustered when the
average connectivity is $\overline l> l_d$, and it disappears for $\overline l>
l_s$. These two thresholds can be computed efficiently using population
dynamics, and in the case of some symmetric problems the value of $l_s$ can be confirmed
rigorously. At the same time, the best algorithms known for random CSP fail to
find solutions in the clustered phase $l_d<\overline l< l_s$. 
This difficulty is due to the `locked' nature of the problem which reduces the clusters
to single points. 
It will be interesting to investigate if LOPs might be used to design new efficient nonlinear error correcting codes, or if the planted LOPs are good candidates for one-way functions in cryptography.  

\paragraph{Acknowledgment --} We thank F. Krzakala, T. Mora and G. Semerjian for many fruitful discussions.

\vspace{-0.3cm}

\bibliography{myentries}

\end{document}